\documentclass[journal]{IEEEtran}

\makeatletter

\newcommand{\Rmnum}[1]{\expandafter\@slowromancap\romannumeral #1@}
\makeatother
\usepackage{graphicx,cite,epsfig,amssymb,amsmath,amsfonts,multirow}
\usepackage{mathrsfs}
\usepackage{float}
\usepackage{soul}
\usepackage{colortbl, color}
\usepackage{cases}
\usepackage[ruled,vlined,linesnumbered]{algorithm2e}
\usepackage{verbatim}

\usepackage{ulem}
\newcommand{\ls}[1]  
   {\dimen0=\fontdimen6\the=#1\dimen0
    \advance\lineskip.5\fontdimen5\the\lineskip-\dimen0
    \lineskiplimit=.9\lineskip
    \baselineskip=\lineskip
    \advance\baselineskip\dimen0
    \normallineskip\lineskip
    \normallineskiplimit\lineskiplimit
    \normalbaselineskip\baselineskip
    \ignorespaces
   }

\begin{document}
\bibliographystyle{ieeetr}

\title{Harvest the Potential of Massive MIMO with Multi-Layer Techniques}


\author{Mingjie~Feng,~\IEEEmembership{Student~Member,~IEEE}~and~Shiwen~Mao,~\IEEEmembership{Senior~Member,~IEEE}%
\thanks{Manuscript received Sept. 3, 2015; revised Nov. 13, 2015; accepted Jan. 4, 2016.
This work was supported in part by the US National Science Foundation under Grants CNS-1247955 and CNS-1320664, and through the Wireless Engineering Research and Engineering Center (WEREC) at Auburn University.}
\thanks{M. Feng and S. Mao are with the Department of Electrical and Computer Engineering, Auburn University, Auburn, AL 36849-5201 USA.
Email: mzf0022@tigermail.auburn.edu, smao@ieee.org.
}
}

\maketitle

\pagestyle{plain}\thispagestyle{plain}

\begin{abstract}
Massive MIMO is envisioned as a promising technology for 5G wireless networks due to its high potential to improve both spectral and energy efficiency. Although the massive MIMO system is based on innovations in the physical layer, the upper layer techniques also play important roles on harvesting the performance gains of massive MIMO. In this paper, we begin with an analysis of the benefits and challenges of massive MIMO systems. We then investigate the multi-layer techniques for incorporating massive MIMO in several important network deployment scenarios. We conclude this paper with a discussion of open and potential problems for future research.
\end{abstract}


\section{Introduction}

The development of the mobile Internet has triggered a fast growing demand for wireless services with high date rates. In the presence of spectrum scarcity, the multiple-input multiple output (MIMO) technology, which offers significant improvements on system capacity and link reliability, is widely acknowledged as a key technology for future wireless systems. In a point-to-point MIMO system with $n_t$ transmitting antennas and $n_r$ receiving antennas, both link reliability and data rate can be improved since the signal is transmitted over ${n_t} \times {n_r}$ channels, with the well-known diversity-multiplexing tradeoff. However, the performance of a point-to-point MIMO system largely depends on the propagation environment. Besides, multiple antennas are required at user terminals (UE), resulting in the increased cost and energy consumption. To overcome these problems and reap the benefits of MIMO systems, the multiple-user MIMO (MU-MIMO) system was developed, where a base station (BS) with multiple antennas serves a set of single-antenna UEs using the same time-frequency resource. Compared to the point-to-point MIMO, the MU-MIMO system is more robust to the propagation environment, while the resulting multi-user diversity gain improves the network-wide performance.

Recently, massive MIMO (also called as large-scale antenna systems, large-scale MIMO, and very large MIMO) has been proposed as a promising technology to dramatically beef up the capacity of wireless networks~\cite{Marzetta10}. A massive MIMO system is characterized by a BS equipped with more than, say, 100 antennas that simultaneously serves multiple users with the same time-frequency resource, where each antenna operates with extremely low power. While preserving all the merits of MU-MIMO, massive MIMO bears some special favorable features due to the law of large numbers. When the number of antennas is large enough, the interference between different users is averaged out, and the effect of fast fading vanishes. Through aggressive spatial multiplexing, both energy and spectral efficiency can be significantly improved since the radio waves sent by the antennas can be controlled so that the energy of the beam can focus on a small region where the intended terminal is located.

Due to the promising perspective, the massive MIMO has been recognized as a key technology in 5G network. Meanwhile, various network architectures were proposed to enhance the capacity of 5G network, such as heterogeneous network, millimeter wave (mmWave) network, device to device (D2D) communication network, cloud radio access network (C-RAN), etc. Thus, as a massive MIMO system is expected to operate under different network scenarios, it is necessary to investigate both the physical layer and upper layer issues to fully harvest the potential of massive MIMO.

In this article, we first review the advantages and technical challenges of massive MIMO. We then present and analyze the multi-layers techniques that are needed to harvest the potential of massive MIMO systems in different network scenarios. Finally, we discuss future research directions and conclude this article.

\section{Advantages and Technical Challenges}

\subsection{Advantages}

Besides considerable performance gains, some key advantages of massive MIMO are summarized in the following.
\begin{itemize}
\item {\em Low Power and Cheap Components}: With massive MIMO, each antenna is transmitting with an extremely low power, in the order of milliwatts. As a result, the requirements for power amplifiers become much less stringent; the system can operate with low-cost amplifier components.
\item {\em Simple Signal Processing Techniques}: For a large number of antennas, the channels between the antennas and different terminals tend to be uncorrelated. Thus, simple precoding and combining schemes, such as maximum ratio combining (MRC) and maximum ratio transmission (MRT), can offer near-optimal performance~\cite{Marzetta10}.
\item {\em Large Number of Degrees of Freedom}: Due to the law of large numbers, the channel response vectors of different terminals become asymptotically-orthogonal, the theoretical number of independent data streams that can be supported equals to the number of antennas. Thus, in a massive MIMO system serving $K$ users with $M$ antennas, the unused degrees of freedom (DoF) is $M-K$~\cite{Larsson14}. The excess DoFs can be used to support more transmissions using the same time-frequency resource. With interference mitigation techniques in the spatial domain such as zero-forcing beamforming, the signals of different links can span on orthogonal subspaces. Such large number of DoF can also be employed to shape the transmitted signal with low peak-to-average power ratio~\cite{Larsson14}.
\item {\em Extremely Robust to Antennas Failures}: With a large number of antennas, the impact of individual antenna failure becomes negligible. The system maintenance cost can thus be reduced.
\end{itemize}

\subsection{Technical Challenges}

The massive MIMO system faces several critical challenges, which are summarized in the following.

\subsubsection{Accurate Channel Estimation with Low Complexity}

To perform efficient detection and precoding, the BS must acquire accurate channel state information (CSI) through channel estimation. In a conventional MU-MIMO system, the BS first transmits pilots to all the UEs; the UEs then estimate the channel and feedback CSI to the BS. However, such process may not be feasible for massive MIMO, since the time devoted to transmit the pilot symbols is proportional to the number of BS antennas; the time spent on channel estimation could be prohibitively long. To avoid transmitting pilots in the downlink, the time division duplex (TDD) system that makes use of channel reciprocity is considered in most literatures. However, a large proportion of current cellular systems are based on the frequency division duplex (FDD) operation, effective solutions are needed to reduce the channel estimation overhead in FDD systems.

\subsubsection{Pilot Contamination}

As discussed in~\cite{Marzetta10}, when the number of users exceeds the number of orthogonal pilot sequences, non-orthogonal pilot sequences have to be used in different cells. In a noncooperative cellular system, the pilot received by a BS is contaminated by transmissions of UEs in other cells that reuse the same pilot. With the contaminated pilots, the beamforming signal causes interference to UEs that share the same pilot.

\subsubsection{Operation with Limited Feedback}

Due to the large number of antennas, it is difficult to acquire instantaneous full CSI at the BS; the system must be able to operate with limited CSI. With limited feedback capability, one way is to design efficient precoding schemes based on partial CSI. The other way is to compress the CSI, then the BS can estimate the full CSI from the compressed feedback. However, more efforts are required to apply these methods to a massive MIMO system due to the large dimension of channel vectors.

\subsubsection{Compatibility with Other Techniques}

The future 5G wireless network is expected to operate with a combination of different techniques, such as small cells, relays, mmWave communications, massive MIMO, D2D communications~\cite{Nishiyama14}, full-duplex transmissions, cloud computing, etc. However, all these techniques need to operate under certain conditions and bear different disadvantages. Thus, new challenges arise when we try to integrate different techniques in an operating 5G cellular system. As an example, when massive MIMO is applied in mmWave communications, the system may operate in short-range scenarios, e.g., a femtocell or picocell with small coverage area, due to the high path loss. Hence the channels of different users may be highly correlated, and user scheduling becomes indispensable~\cite{Swindlehurst14}.

\subsubsection{Energy Efficiency}

Although massive MIMO is energy efficient is terms of signal transmission, other means of power consumption such as the circuit power potentially degrades the
system energy efficiency. In the presence of massive arrays, vast amount of data would be generated and processed. Without efficient baseband signal processing
schemes, the internal power consumption would be much higher than expected.

\section{Multi-Layer Techniques in Different Network Scenarios with Massive MIMO}

\subsection{Homogeneous Cellular Networks}

\subsubsection{Single-cell Scenario}

The resource allocation within a single cell was considered in~\cite{Ng12}, where a joint subcarrier allocation, power allocation, antenna allocation and data rate adaptation scheme was proposed to improve the energy efficiency. In this model, the power of the amplifier is determined by subcarrier selection and the power on each subcarrier, while the circuit power of antennas is proportional to the number of active antennas. Thus, for the sake of energy saving, not all antennas are activated for transmission and the BS does not make use of all the subcarriers. The problem was formulated as a nonlinear fractional programming, and solved with an iterative algorithm.

To reduce the channel estimation overhead, an efficient approach is user scheduling strategy, such as user grouping. In a typical massive MIMO system that supports user grouping, the precoding process is divided into two stages. In the first stage, users with similar second-order channel statistics are put into one group, the same precoding is used for users in the same group. Then, a second stage dynamic precoding is applied with the reduced feedback by user grouping. The performances of different user grouping strategies, such as K-means, weighted likelihood, subspace projection, fubini study, hierarchical based, and K-medoids were investigated in~\cite{Xu14}. Furthermore, a user scheduling algorithm was proposed in~\cite{Xu14} that selects the sets of users to transmit in each time slot.

\subsubsection{Multi-cell Scenario}

\begin{figure}[!t]
 \begin{center}
   \includegraphics[width=3.4in]{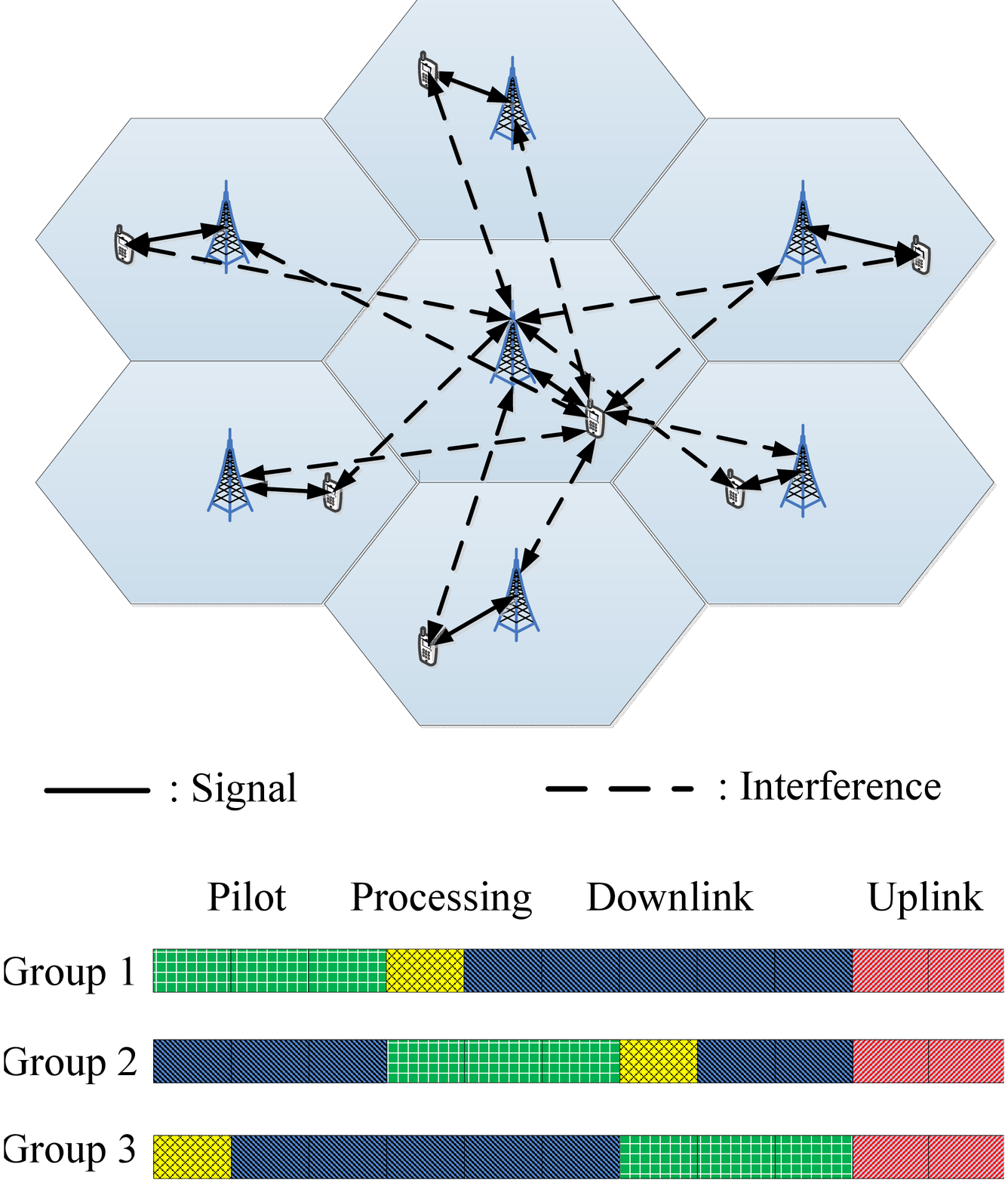}
 \end{center}
\caption{An example of a time-shifted frame structure for pilot contamination mitigation \cite{Fernandes13}.}
\label{fig1}
\end{figure}

In a noncooperative cellular network, the interference from uncorrelated links can be averaged out with linear processing, the pilot contamination becomes the only factor that impacts the signal-to-interference ratio (SIR)~\cite{Marzetta10}. Apart from various signal processing techniques, pilot contamination can also be mitigated with upper layer approach such as pilot allocation. In a multi-cell network, the reuse pattern of pilots can be optimized so that UEs with low inter-cell interference are allocated with the same pilot. Another efficient way to mitigate pilot contamination is to design the MAC protocol of transmission frames. In~\cite{Fernandes13}, a time-shifted frame structure was proposed, in which all the cells are divided into several groups, and cells in these groups transmit their pilots at different time slots, as shown in Fig.~\ref{fig1}. Thus, the aggregated interference caused by other cells is mitigated, resulting in good performance gain.

\subsection{Heterogeneous Cellular Networks}

In a heterogeneous network with massive MIMO, the law of large numbers holds for the following two cases. The first one is when joint precoding and joint combining are performed between macrocell and small cells. In this way, the antennas at small cell BS (SBS) become part of the massive MIMO array, the interference between uncorrelated links can be averaged out. However, a large overhead would generate between macrocell BS (MBS) and SBS, and each SBS needs to detect and estimate the signals of all users. Thus, the design and application of such a system would be quite challenging. For the second case, the MBS estimates the channels of all the SBS's, small-cell UEs (SUE), and macrocell UEs (MUE), and then apply linear combining and precoding to detect and transmit the signals of MUEs. With this approach, the SBS's and SUEs are regarded as virtual users to be served, hence the interference between small cell and macrocell transmissions can be averaged out.

\subsubsection{Interference Management}

\begin{figure}[!t]
 \begin{center}
   \includegraphics[width=3.4in]{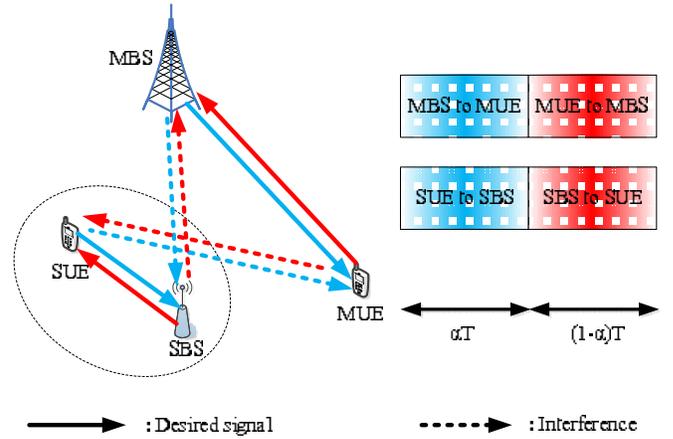}
 \end{center}
\caption{Reversed TDD in a heterogeneous cellular network.}
\label{fig2}
\end{figure}

Interference management in a heterogeneous network with massive MIMO would be challenging due to the increased difficulty to acquire CSI for all antennas. In~\cite{Hosseini13}, an interference management scheme was proposed, where the uplink and downlink transmissions of the two tiers are performed in a reversed pattern as shown in Fig.~\ref{fig2}, referred to as reverse TDD. With this approach, the cross-tier interference occurs between MBS and SBS, and between MUE and SUE. Since the transmit powers of MUEs and SUEs are relatively lower, the mutual interference can be maintained at a low level. To handle the interference between MBS and SBS's, the MBS first estimates the interference from SBS's by subtracting the desired signal from the received signal. Then, the MBS performs precoding by projecting the precoding vectors to the subspace orthogonal to the subspace of SBS's interfering signals, so that the interference between MBS and SBS's is eliminated. Due to the large number of DoFs, the performance loss of MUEs resulted from interference avoidance can be kept small.

\subsubsection{User Association}

As both massive MIMO and small cells have their advantages and limitations, the choices of user association need to take several factors into consideration, such as channel condition, interference pattern, traffic loads at MBS and SBS's, and backhaul capacity, etc. Consider the load balancing between MBS and SBS's as an example. On one hand, when more users are connected to an SBS, the resource for each user decreases, and the increased traffic load brings pressure to the backhaul. On the other hand, as more users are connected to the MBS with massive MIMO, the time spent on channel estimation is increased, resulting in decreased average throughput for users served by MBS.

In~\cite{Xu15}, user association in a heterogeneous network with massive MIMO was formulated as an integer programming problem with objective of rate maximization. Since the formulated integer programming problem is NP-hard, it was relaxed to a linear programming problem, and the solution of the relaxed problem was proven to be optimal for the original problem. Moreover, two distributed schemes were proposed to reduce the overhead, one is the based on service provider pricing, the other is based on user bidding. Both the pricing and bidding games were proven to converge to a Nash Equilibrium (NE).

The model in~\cite{Xu15} assumes that each user can only connect to one BS. With the aid of coordinated multi-point transmission, a user can be served by multiple BS's. Under this assumption, user association and beamforming were jointly considered to improve system energy efficiency in~\cite{Bjornson13}. The problem was formulated with the objective to minimize the energy consumption, subject to qualify of service (QoS) constraints for users. Due to the hidden convexity, the formulated problem can be transformed with a semi-definite relaxation without loss of optimality, and the optimal solution can be obtained.

\subsection{Relay Networks}

An advantage of applying massive MIMO in relay networks is the effective reduction of loop interference, which enables a full-duplex transmission pattern. When the relay station is equipped with a large number of transmit and receive antennas, the transmitted signal can be project to the subspace that is orthogonal to the received signal, resulting in reduced loop interference. It was demonstrated in~\cite{Ngo14} that the interference between different transmit-receive pairs vanishes and noise effects disappear as the numbers of receive and transmit antennas, ${N_{rx}}$ and $N_{tx}$ are large enough. Besides, the transmit powers of each source and of the relay proportionally to $\frac{1}{{{N_{tx}}}}$ and $\frac{1}{{{N_{rx}}}}$, respectively. The achievable rates with zero-forcing (ZF) and MRT/MRC processing were derived in~\cite{Ngo14}, with an optimal power allocation algorithm to improve the system energy efficiency.

Note that, there is only one relay station in the model of~\cite{Ngo14}. In a relay network with multiple relay stations, coordination among relay stations may be required for interference management and performance enhancement.

\section{Future Research Directions}

\subsection{Heterogeneous Networks}

\subsubsection{Wireless Backhaul with Massive MIMO}

The backhaul data transmission between MBS and SBS has two options, a wireline connection or a wireless connection. Although a wired backhaul is more stable and supports high data rate, a wireless backhaul is desirable in terms of easy implementation, flexible configuration and low cost. Moreover, when the MBS is equipped with a large number of antennas, a high data rate is achievable between MBS and SBS through the wireless backhaul.

\begin{figure}[!t]
 \begin{center}
   \includegraphics[width=3.4in]{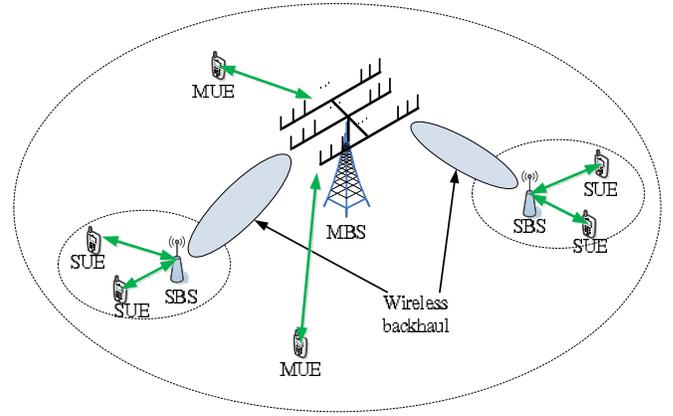}
 \end{center}
\caption{Wireless backhaul in a heterogeneous network with massive MIMO.}
\label{fig3}
\end{figure}

As shown in Fig.~\ref{fig3}, three kinds of wireless transmissions coexist in the two-tier heterogeneous network with wireless backhaul, resulting in a more complicated interference pattern. Thus, the MAC layer schedules such as spectrum, time, and power allocations need to be investigated. Another issue is the choice between TDD and FDD, which should be jointly considered with other scheduling factors. Besides, the inherent challenges of massive MIMO systems, e.g., pilot contamination, low-complexity channel estimation, and operation under limited feedback, should also be carefully addressed under the wireless backhaul scenario.

\subsubsection{Offloading Strategy}

The deployment of small cells helps to offload the MBS traffic, resulting in reduced overhead and congestion at MBS. However, with the limited spectrum and power resource of each SBS and the limited backhaul capacity, the offloading capability of small cells is limited. Thus, the offloading strategy should consider various factors such as channel condition of each user, available resources at the BS's, instantaneous traffic condition, etc. To reduce congestion and improve QoS, load balancing would be desirable so that traffic load can be switched between the SBS's. A possible approach for load balancing is to adjust the sizes of small cells, such that the number of active users in each small cell can be controlled.

\subsubsection{Hotspot Coverage}

To serve a large number of users in a hotspot, the wireless network has to be densified to improve spatial spectrum reuse. Although the small cell has been recognized as an effective approach for hotspot coverage, future wireless networks would be challenged by the extremely high data requirements in hotspots. Massive MIMO has the potential to deal with this challenge since it allows aggressive spatial multiplexing to improve network capacity, thus, a combination of small cells and massive MIMO may be considered in future wireless network architectures. Specifically, we can deploy SBS's with a large antenna array to serve users in hotspots. To equip a large number of antennas in an SBS with small size, the antennas have to be densely placed, the distance between neighboring antennas can be in the order of centimeter or millimeter. Thus, the system may operate in higher frequency spectrum bands with large propagation loss. Although such attenuation can be mitigated by the array gain brought by massive MIMO, the coverage area of an SBS is still relatively small, multiple SBS's are required to serve a hotspot. Thus, problems such as interference management, resource allocation, load balancing, and cooperation of small cells need to be investigated. Moreover, in a hotspot where users are close to each other (e.g. in a football stadium or a commercial center), and there is no rich scattering environment, the channels of different UEs may be correlated. Hence, the scaling law of linear processing does not hold, the interference between different users can not be averaged out, and more sophisticated schemes are required to deal with such interference.

\subsection{MmWave Network with Massive MIMO}

The combination of mmWave and massive MIMO was investigated in~\cite{Swindlehurst14}. Apart from the challenges on physical layer techniques and hardware designs, interference management was mentioned as an issue that requires further study. Although mmWave transmissions can be viewed as pseudo-wired when the beam is extremely narrow, the co-channel interference resulted from scattering and reflection between different links becomes a rising concern in recent literature. In a mmWave network with overlapping beams and reuse the same frequency, an effective approach to avoid interference is link schedule. For example, the links with mutual interference can be scheduled to transmit at different time instants. With massive MIMO, the interference mitigation can also be performed in the space domain using excess DoFs.

For an indoor mmWave network with massive MIMO, due to the lack of rich scattering environment, the channel orthogonality may not hold for all users. Thus, the scaling law does not hold, it is necessary to carry out interference mitigation schemes such as user scheduling and adaptive beamforming.

\begin{figure}[!t]
 \begin{center}
   \includegraphics[width=3.0in]{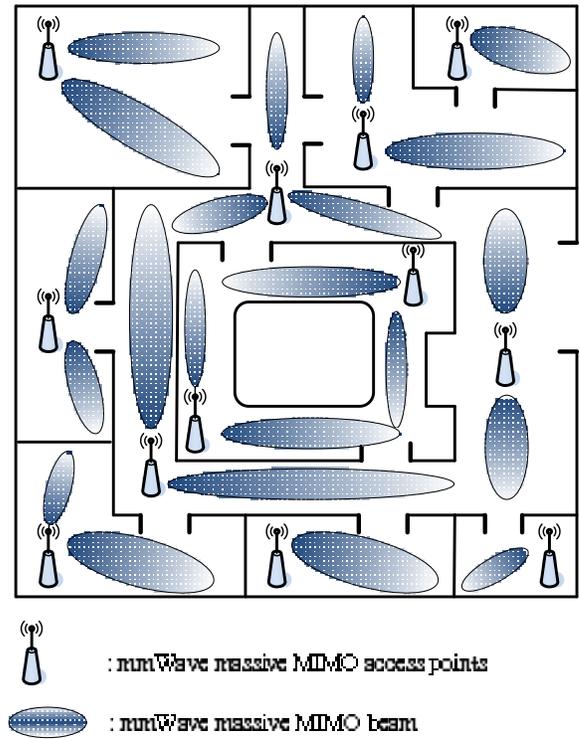}
 \end{center}
\caption{An example of mmWave massive MIMO application scenario.}
\label{fig4}
\end{figure}

An example of indoor mmWave network is shown in Fig.~\ref{fig4}. Since the mmWave signals can not penetrate obstacles like walls, the signals can only propagate via line of sight (LOS) transmissions or reflection. As a result, the locations and transmission patterns of the access points have to be carefully scheduled to guarantee the coverage performance as shown in Fig.~\ref{fig4}. Moreover, other network planning issues including coverage area adjustment, resource allocation, and user handover are more challenging than existing wireless networks due to the transmission restrictions of mmWave signals. Thus, highly adaptive network schedule strategies are necessary for deployment of mmWave massive MIMO networks.

\subsection{Massive MIMO with Cloud Computing}

\begin{figure}[!t]
 \begin{center}
   \includegraphics[width=3.4in]{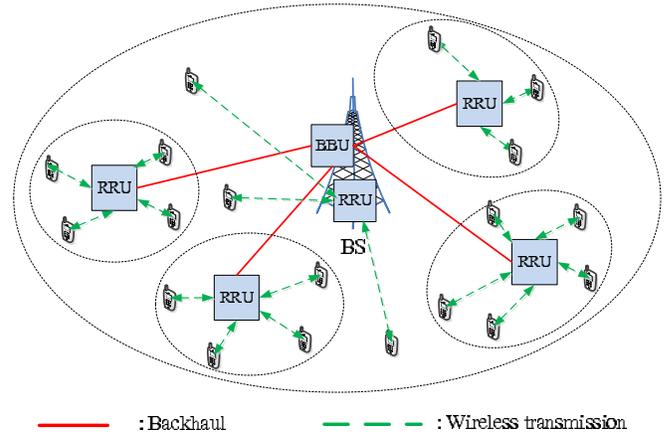}
 \end{center}
\caption{Massive MIMO in a cellular network with distributed RRUs.}
\label{fig5}
\end{figure}

The large number of antennas generate huge amount of data at the BS. Although massive MIMO allows simple signal processing techniques such as MRC and MRT, the large dimension may still cause high complexity of data processing at the BS. A potential approach to deal with this challenge is to offload data processing to the cloud, so that the high complexity computations can be executed in the cloud (as in the emerging cloud radio access network (C-RAN) architecture). As shown in Fig.~\ref{fig5}, distributed antenna arrays are deployed at remote radio units (RRU). The RRUs are connected to the baseband units (BBU) at the BS through a backhaul. With this architecture, a large number of antennas can be placed in more than one sites to extend coverage and enhance throughput, each RRU is connected to part of the antennas. In~\cite{Li14}, a massive MIMO enabled C-RAN model was considered where a large number of remote radio heads (RRH) were deployed. Specifically, the transmission power of each RRH cluster is optimized to maximize the network energy efficiency.

The model in~\cite{Li14} can be viewed as a distributed massive MIMO system based on C-RAN architecture. The cloud computing can also be applied for a system with co-located antennas in one site. It is worth mentioning that offloading to the cloud comes at a price, the data transmission between BS and cloud requires additional communication expenditure and results in additional latency. Thus, the choice between cloud offloading and local execution should be carefully decided. Besides, scheduling issues including resource allocation, task assignment, surrogate selection, and load balancing need to be studied to enhance the system performance in terms of latency and energy efficiency.

\subsection{Green Massive MIMO Systems}

\begin{figure}[!t]
 \begin{center}
   \includegraphics[width=3.0in]{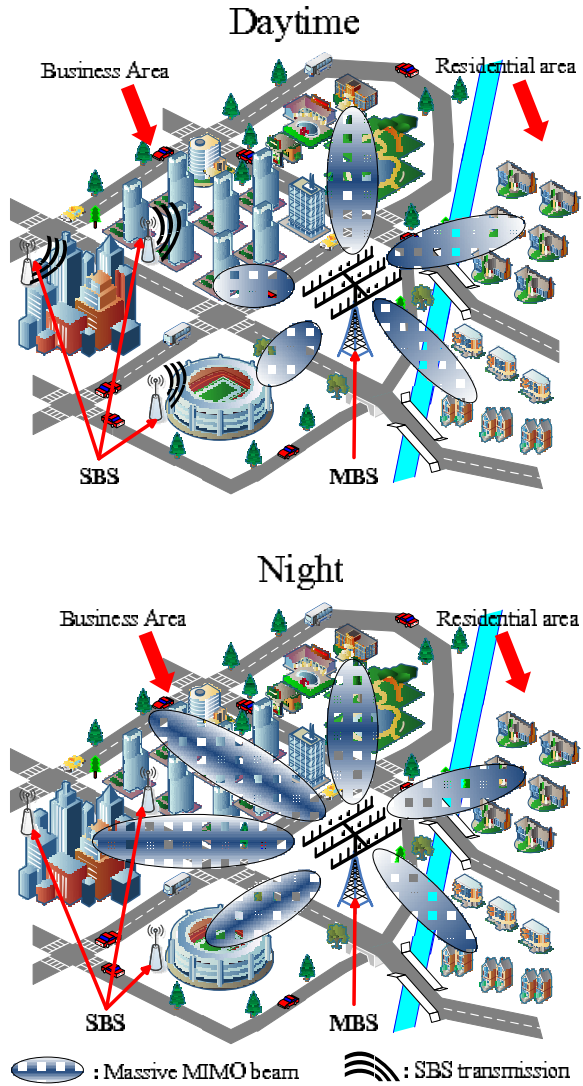}
 \end{center}
\caption{Illustration of energy efficiency improvement in a heterogeneous network with massive MIMO through BS ON/OFF schedule.}
\label{fig6}
\end{figure}

Although massive MIMO is highly energy efficient for signal transmissions, the energy consumption of hardware components should be considered to improve energy efficiency. In~\cite{Ng12}, the antenna selection problem was considered where the set of active antennas was chosen to maximize energy efficiency. The application of electromagnetic lens in a massive MIMO system was investigated in~\cite{Zeng14}. Due to the advantages energy focusing and spatial interference rejection, the number of required radio frequency (RF) chains can be significantly reduced through low complexity antenna selection schemes, resulting in reduced energy consumption. In a heterogeneous network with massive MIMO, the SBS's can be dynamically turned on and off according to traffic dynamics to save energy, while the coverage can be compensated by the MBS with massive MIMO~\cite{Feng16}. As shown in Fig.~\ref{fig6}, when the traffic load of an SBS is reduced, e.g., an SBS located at a shopping mall at nighttime, the SBS can be turned off to save energy, and the MBS with massive MIMO would provide coverage to the area. Such a strategy is favorable for a massive MIMO system since the coverage area of an MBS can be well adjusted due to the excellent spatial focus.

\subsection{Cooperative Massive MIMO}

In~\cite{Huh12}, a cooperative network MIMO architecture was proposed. In this work, the BS's are divided into clusters to enable a cooperative network-MIMO scheme, including designs on linear zero-forcing beamforming with suitable inter-cluster interference constraints, uplink pilot signals allocation and frequency reuse across cells. Users are partitioned into classes based on geographical locations, and users in the same class are served with same time-frequency using a network-MIMO scheme that is specifically optimized for the class. The proposed network-MIMO architecture was shown to achieve comparable spectral efficiency with a massive MIMO system using much fewer number of antennas.

In~\cite{Hosseini14}, a cooperative network MIMO system was considered and compared with a massive MIMO system. In the network MIMO system, the BS's perform joint transmission through data and CSI exchange with backhaul links, yielding a cooperative transmission pattern. Although it was demonstrated in~\cite{Hosseini14} that massive MIMO outperforms network MIMO in terms of interference mitigation, other means of inter-cell cooperation can be employed other than joint transmission. For example, the BS's can cooperate for load balancing, power control, etc. Besides, the cooperation between BS's with massive MIMO is a potential approach to further enhance the performance while it has not been well studied yet. However, the cooperation between massive MIMO BS's faces the challenge brought by large amount of data generated at each BS; thus, efficient cooperation approaches with limited overhead are required to harvest the potential.

\subsection{Full-Duplex Massive MIMO}

The full-duplex transmission is a promising approach to improve spectral efficiency by allowing a transceiver to transmit and receive simultaneously on the same frequency. The full-duplex relay network with massive MIMO was considered in~\cite{Ngo14}, while the application of full-duplex transmissions in massive MIMO cellular networks has not been investigated yet. In a full-duplex massive MIMO cellular system, the same frequency band can be simultaneously reused by two groups of users, with one group used for uplink and the other group used for downlink, thus improving the spectrum utilization. The large number of antennas at BS may be advantageous for self-interference cancellation, but technical details such as user scheduling, antennas selection would be critical for the successful application in future wireless networks.

\section{Conclusion}

As the massive MIMO is expected to operate with other techniques in 5G network, multi-layers issues need to be considered to harvest its potential. In this article, we described and analyzed multi-layer techniques in massive MIMO systems with different network scenarios. We also discussed the potential research topics for future research in order to fully harvest the high potential of massive MIMO.


\vfil

\begin{biography}
{Mingjie Feng} [S'15] received his B.E. and M.E. degrees from Huazhong University of Science and Technology in 2010 and 2013, respectively, both in electrical engineering. He was a visiting student in the Department of Computer Science, Hong Kong University of Science and Technology, in 2013. He is currently a Ph.D. student in the Department of Electrical and Computer Engineering, Auburn University, Alabama. His research interests include cognitive radio networks, femtocell networks, massive MIMO and full-duplex communication. He is a recipient of a Woltosz Fellowship at Auburn University.
\end{biography}

\vfil

\begin{biography}
{Shiwen Mao} [S'99-M'04-SM'09] received Ph.D. in electrical and computer engineering from Polytechnic University, Brooklyn, NY. He is the Samuel Ginn Distinguished Professor and Director of the Wireless Engineering Research and Education Center (WEREC) at Auburn University, Auburn, AL. His research interests include wireless networks and multimedia communications. He is a Distinguished Lecturer of the IEEE Vehicular Technology Society (VTS) in the Class of 2014. He is on the Editorial Board of IEEE Transactions on Multimedia, IEEE Internet of Things Journal, IEEE Communications Surveys and Tutorials, and IEEE Multimedia, among others. He received the 2015 IEEE ComSoc TC-CSR Distinguished Service Award, the 2013 IEEE ComSoc MMTC Outstanding Leadership Award, and the NSF CAREER Award in 2010. He is a co-recipient of the Best Paper Awards from IEEE GLOBECOM 2015, IEEE WCNC 2015, and IEEE ICC 2013, and the 2004 IEEE Communications Society Leonard G. Abraham Prize in the Field of Communications Systems.
\end{biography}

\vfil

\end{document}